\documentclass[5p]{elsarticle}
\biboptions{authoryear}

\usepackage{amssymb}
\usepackage{latexsym}

\usepackage{graphicx}
\usepackage{glossaries}
\usepackage{algorithm}
\usepackage{algorithmic}
\usepackage{amsthm}

\usepackage{amsmath}  
\usepackage{subfigure}
\usepackage{bbm}
\usepackage{multirow}
\usepackage{booktabs}
\usepackage{float}

\usepackage{fontawesome}
\floatname{algorithm}{Algorithm}

\usepackage{url}
\usepackage{xcolor}

\usepackage[colorlinks]{hyperref}

\definecolor{newcolor}{rgb}{.8,.349,.1}

\makeatletter
\def\ps@pprintTitle{%
   \let\@oddhead\@empty
   \let\@evenhead\@empty
   \let\@oddfoot\@empty
   \let\@evenfoot\@oddfoot
}
\makeatother

\begin{document}
\begin{frontmatter}
\title{Federated Learning for Computational Pathology on Gigapixel Whole Slide Images}%
\author[1,3]{Ming Y. Lu}
\author[1]{Dehan Kong}
\author[1,3]{Jana Lipkova}
\author[1,2,3]{Richard J. Chen}
\author[5]{Rajendra Singh}
\author[1,3]{\\Drew F. K. Williamson}
\author[1,3]{Tiffany Y. Chen}
\author[1,3,4]{Faisal Mahmood$^*$}
\ead{faisalmahmood@bwh.harvard.edu}
\cortext[cor1]{Corresponding author:}
\address[1]{Department of Pathology, Brigham and Women's Hospital, Harvard Medical School, Boston, MA}
\address[2]{Department of Biomedical Informatics, Harvard Medical School, Boston, MA}
\address[3]{Cancer Program, Broad Institute of Harvard and MIT, Cambridge, MA}
\address[4]{Cancer Data Science, Dana-Farber Cancer Institute, Boston, MA}
\address[5]{Department of Pathology, Northwell Health, NY}

\begin{abstract}
Deep Learning-based computational pathology algorithms have demonstrated profound ability to excel in a wide array of tasks that range from characterization of well known morphological phenotypes to predicting non human-identifiable features from histology such as molecular alterations. However, the development of robust, adaptable and accurate deep learning-based models often rely on the collection and time-costly curation large high-quality annotated training data that should ideally come from diverse sources and patient populations to cater for the heterogeneity that exists in such datasets. Multi-centric and collaborative integration of medical data across multiple institutions can naturally help overcome this challenge and boost the model performance but is limited by privacy concerns amongst other difficulties that may arise in the complex data sharing process as models scale towards using hundreds of thousands of gigapixel whole slide images. In this paper, we introduce privacy preserving federated learning for gigapixel whole slide images in computational pathology using weakly-supervised attention multiple instance learning and differential privacy. We evaluated our approach on two different diagnostic problems using thousands of histology whole slide images with only slide-level labels. Additionally, we present a weakly-supervised learning framework for survival prediction and patient stratification from whole slide images and demonstrate its effectiveness in a federated setting. Our results show that using federated learning, we can effectively develop accurate weakly supervised deep learning models from distributed data silos without direct data sharing and its associated complexities, while also preserving differential privacy using randomized noise generation.
\end{abstract}

\end{frontmatter}


\section{Introduction}
\begin{figure*}[ht]
\includegraphics[width=\textwidth]{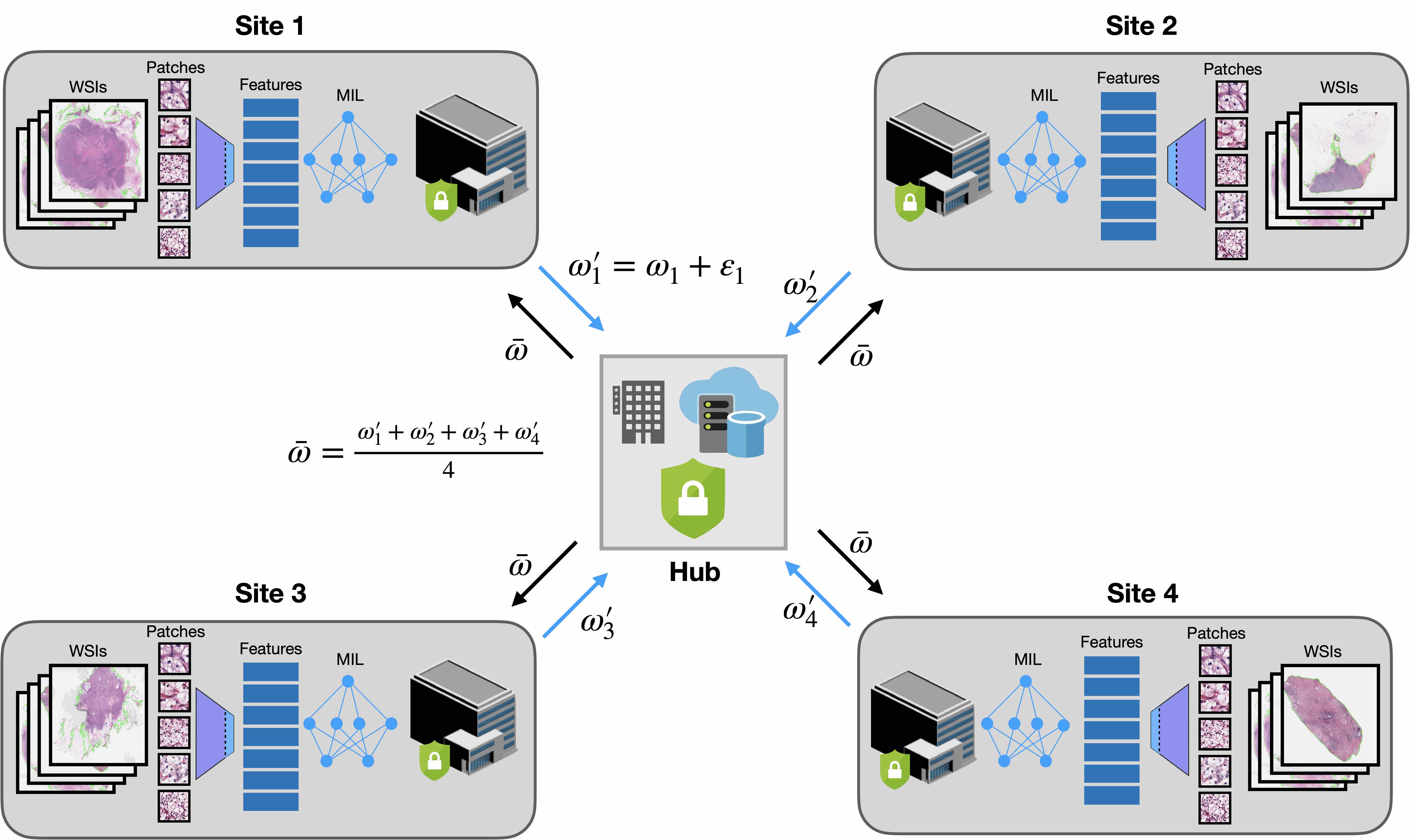}
\caption{\textbf{Overview of the weakly supervised multiple instance learning in a federated learning framework.} At each client site, for each WSI, the tissue regions are first automatically segmented and image patches are extracted from the segmented foreground regions. Then all patches are embedded into a low-dimension feature representation using a pretrained CNN as the encoder. Each client site trains a model using weakly-supervised learning on local data (requires only the slide-level or patient-level labels) and sends the model weights each epoch to a central server. Random noise can be added to the weight parameters before communicating with the central hub for differential privacy preservation. On the central server, the global model is updated by averaging the model weights retrieved from all client sites. After the federated averaging, the updated weights of the global model is then sent to each client model for synchronization prior to starting the next federated round.}
\end{figure*}

The emerging field of computational pathology holds great potential in increasing objectivity and enhancing precision of histopathological examination of tissue.
Machine learning -- and deep learning in particular -- have demonstrated unprecedented performance in various pathology tasks such as characterization of a disease phenotype \citep{wang2020weakly,zhou2019cgc,anand2020histographs,bulten2020automated,mahmood2019deep}, quantification of the tumor microenvironment  \citep{javed2020cellular,graham2019hover,schapiro2017histocat}, prediction of survival  \citep{muhammad2019unsupervised} and treatment response \citep{niazi2019digital,bera2019artificial}, and integration of genomics with histology for improved patient stratification  \citep{chen2019pathomic,mobadersany2018predicting,lazar2017comprehensive}. Thanks to the ability of such algorithms to mine sub-visual features -- even beyond the scope of known pathological markers -- deep learning models have managed to tackle challenging tasks such as estimating primary source for metastatic tumors of unknown origin \citep{lu2020deep}, identifying novel features of prognostic relevance \citep{yamamoto2019automated,pell2019use,bera2019artificial},  and predicting genetic mutations from histomorphologic images only, without the use of immunohistochemical staining \citep{coudray2018classification}. Among various approaches, weakly-supervised methods such as attention MIL \citep{lu2019semi,lu2020data} appear well-suited to potential adoption in clinical practice. These models learn from weak annotations in the form of \textit{image or patient-level labels} which can include labels such as diagnosis or survival associated with the patient. Such information is readily available in clinical records and thus the data annotation does not introduce significant overhead over standard clinical workflow, in contrast to \textit{pixel-level} annotations of regions of interest required by supervised models. 
\\
As in all machine learning affairs, the model's accuracy and robustness can be significantly increased by incorporating diverse data reflecting variations in underlying patient populations as well as data collection and preparation protocols. Specifically, in pathology, the whole slide images (WSIs) used for computational analysis can exhibit immense heterogeneity which arises from not only the patient group corresponding the histology specimens and variations in the tissue preparation, fixation and staining protocols but also different scanner hardware that are used for digitization.
While it may be possible and desirable to gain increased exposure to such heterogeneity through agglomeration of medical data from multiple institutions into a centralized data repository in order to develop more generalizable models, data centralization poses challenges not only in the form of regulatory and legal concerns (e.g. differences in jurisdictions might prevent data transfer among countries \cite{scheibner2020data}) but also technical difficulties such as high cost of transfer and storage of huge quantities of data. The latter is particularly relevant for computational pathology at scale since just 500 gigapixel WSIs can be as large as the entirety of ImageNet \citep{deng2009imagenet}.
\\
Federated learning  \citep{yang2019federated,konevcny2016federated,mcmahan2017communication,rieke2020future} offers means to mitigate these challenges by enabling algorithms to learn from de-centralized data distributed across various institutions. In this way, sensitive patient data are never transferred beyond the safety of institutional firewalls, and instead, the model's training and validation occur locally at each institution and only the model specifics (e.g. parameters or gradients) are transferred. In general, federated learning can be achieved through two approaches. \textit{1) Master-server}: a master-server is used to transfer the model to each node (i.e. participating institution), where the model trains for several iterations using the local data. The master-server then collects the model parameters from each node, aggregates them in some manner, and updates the parameters of the global model. Updated parameters are then transferred back to the local nodes for the next iteration. \textit{2) Peer-to-peer}: each node transfers the locally-trained parameters to some or all of its peers and each node does its own parameter aggregation. The benefit of the master-server approach is that all governing mechanisms are separated from the local nodes which allows for easier protocol updates and inclusion of new institutions. In contrast, the peer-to-peer approach is less flexible -- since all the protocols must be agreed-on in advance -- however, the absence of a single controlling entity might be preferred in some cases e.g. due to lower costs or greater decentralization.
\\
Although the nodes never transfer data themselves -- only the model specifics -- if leaked or attacked these can be sufficient to indirectly expose sensitive private information. Data anonymization alone does not provide sufficient protection  \citep{rocher2019estimating} since parts of the training data can be reconstructed by inversion of model parameters  \citep{carlini2018secret,zhang2020secret}, gradients  \citep{zhu2019deep}, or through adversarial attacks  \citep{wang2019beyond,hitaj2017deep}. This is particularly worrisome in radiology where the medical scans can be used to reconstruct a patient's face or body image. Even though histology data do not hold such a direct link with patient identity, it might still allow an indirect patient identification e.g. in the case of rare diseases. The design of countermeasures for increasing differential privacy is thus a very active field of research  \citep{kaissis2020secure,kairouz2019advances}. A popular strategy in the medical field is a contamination of the input data  \citep{cheu2019distributed} or the model parameters  \citep{dong2019gaussian} with certain levels of noise. This decreases the individually recognizable information while preserving the global distribution of the data  \citep{kaissis2020secure}. \\
Though federated learning was originally proposed for non-clinical use, since its inception in 2016  \citep{konevcny2016federated}, it has already appeared in some medical applications. These include large-scale multicenter studies of genomics \citep{mandl2020genomics,rehm2017evolving,jagadeesh2017deriving}, electronic health records \citep{brisimi2018federated,choudhury2019differential,choudhury2019predicting}, or wearable health devices  \citep{chen2020fedhealth}. In the field of medical imaging, federated learning is particularly popular in the neurosciences. So far it has been applied in tasks such as brain tumor \citep{li2019privacy,sheller2018multi} and brain tissue \citep{roy2019braintorrent} segmentation, EEG signal classification \citep{ju2020federated}, analysis of fMRI scans of patients with autism \citep{li2020multi} or MRI scans of neurodegenerative disease  \citep{silva2019federated}. Further adoption of federated learning in other medical domains is strongly anticipated due to the increasing demand for large and inter-institutional studies. \\
One of the fields that would strongly benefit from the federated framework is computational pathology. Since histopathologic diagnosis is the gold standard for many diseases, pathology data are largely available in almost any hospital. Federated learning would in principle enable deep learning models to learn from much larger and more diverse multi-institutional data sources without the challenges associated with data centralization. Furthermore, while fully-supervised approaches are burdened by the need for time-costly pixel-level annotation based on pathologist expertise, weakly-supervised approaches such as MIL simplify collaborative efforts by alleviating the requirement for such human expertise and the burden of creating pixel-level labels under unified annotation protocol in all participating institutions. \\

Herin, we present the \textbf{key contributions} of our work as follows: 
\begin{enumerate}
  \item We present the first large-scale computational pathology study to demonstrate the feasibility and effectiveness of privacy-preserving federated learning using thousands of gigapixel whole slide images from multiple institutions. 
  \item We account for the challenges associated with the lack of detailed annotations in most real world whole slide histopathology datasets and are the first to demonstrate how federated learning can be coupled with weakly-supervised multiple instance learning to perform both binary and multi-class classification problems (demonstrated on breast cancer and renal cell cancer histological subtyping) using only slide-level labels for supervision.  
  \item We extend the usage of attention-based pooling in multiple instance learning-based classification and present a weakly-supervised framework for survival prediction (demonstrated on renal cell carcinoma patients) in computational pathology using whole slide images and patient-level prognostic information, without requiring manual ROI-selection or randomly sampling a predetermined number of patches. 
\end{enumerate}

\section{Methods}
In this section, we will formulate our weakly-supervised federated multiple instance learning framework for performing privacy-preserving federated learning on data from across multiple institutions in the form of digitized gigapixel whole slide images. 

\subsection{Differential privacy and federated learning} 
In this problem, we want to develop deep learning models for performing predictive tasks on gigapixel WSIs by using data from different institutions. We denote data owned by institution i as $D_i$, which we assume for simplicity, is simply a data matrix with a finite number of entries. Suppose there are in total B sites and we denote their corresponding data silo as $D_1, D_2, \cdots, D_B$. Since each medical institution will not share data with other parties due to various issues (\textit{e.g.} institutional policies, incompatible data sharing protocols, technical difficulties associated with sharing large amount of data or fear of privacy loss), we cannot pool together their data and train a single deep learning model $\it{f}_{centralized}$ for solving the desired task. Instead, our objective is to develop a federated learning framework where the data owners collaboratively train a model $\it{f}_{global}$, in which each data owner does not need to share its data $D_i$ with others but can all benefit from the usefulness of the final model. In this paper, we adapted a master-server architecture, where each client node, representing each medical institution, locally utilizes the same deep learning architecture as one another and the global model, which we assume to be hosted on a central server hub. Each institution trains its respective model using local data and uploads the values of the trainable model parameters to the master server at a consistent frequency (\textit{i.e.} once every one epoch of local training). We also adopt a randomized mechanism utilized in a previous study \citep{li2020federated} on multi-site fMRI analysis, which allows each data owner to  blur the shared weight parameters by a randomly noise $z_i$ to protect against leakage of patient-specific information. After the master server receives all the parameters, it averages them in the global model and sends new parameters back to each local model for synchronization. \\
Differential privacy is a popular definition of individual privacy \citep{dwork2014algorithmic,shokri2015privacy}, which informally means that the attacker can learn virtually nothing about an individual sample if it were removed from or added to the dataset \citep{abadi2016deep}. In this problem, it means when a data point $d_i$ is removed from or added to the dataset $D_i$, the attacker can not infer any information about $d_i$ from the output or weights of model $\it{f}_{global}$. Differential privacy provides a bound $\epsilon$ to represent the level of privacy preference that each institution can control. Formally, it says \citep{dwork2006our}, given a deterministic function $\it{f}$, and two adjacent datasets $D_i$, $D_i'$ differing by exactly one example, \textit{i.e.}, $\left\|{D_i-D_i'}\right\|_1 = 1$, $\it{f}$ satisfies $(\epsilon,\delta)$-differential privacy if for any subset of outputs $S$:
\begin{equation}
    \it{P}[\it{f}(D_i)\in\it{S}]\leq\it{e^{\epsilon}\it{P}[\it{f}(D_i')\in\it{S}]}+\delta
\end{equation}
where the introduction of the $\delta$ term relaxes the stricter notion of $\epsilon$-differential privacy and allows the unlikely event of differential privacy being broken to occur with a small probability. To satisfy $(\epsilon,\delta)$-differential privacy, first, we provide the defintion of $\it{l_2}$ sensitivity of $\it{f}$, denoted by $\Delta_{2}(f)$, as:
\begin{equation}
\Delta_{2}(f)=\max _{\atop\|D_i-D_i'\|_{1}=1}\|f(D_i)-f(D_i')\|_{2}
\end{equation}
For arbitrary $\epsilon\in(0,1)$, as stated by Theorem 3.22 \citep{dwork2014algorithmic}, adding random noise to \textit{f} that is generated from a Gaussian distribution with zero mean and standard deviation equal to $\Delta_{2}(f)\sigma$, \textit{i.e.}, $\mathcal{N}(0,\Delta_{2}(f)^2\sigma^2)$, will result in \textit{f} satisfying $(\epsilon,\delta)$-differential privacy if $\sigma \geq \frac{c\Delta_{2}(f)}{\epsilon}$ and $c^2>2\ln{\frac{1.25}{\delta}}$. After rewriting the two inequalities, in other words, for any choice of $\epsilon$, $(\epsilon,\delta)$-differential privacy can be satisfied for $\it{f}$ by using the Gaussian mechanism, where $\delta$ is related to the variance of the Gaussian noise distribution via: 
\begin{equation}
\begin{split}
{\sigma}^2 & \geq \frac{c^2\Delta_{2}^2(f)}{\epsilon^2} \implies {\sigma}^2 > \frac{2\ln{(\frac{1.25}{\delta}})\Delta_{2}^2(f)}{\epsilon^2} \\ \frac{\epsilon^2\sigma^2}{2\Delta_{2}(f)^2} & >\ln{(\frac{1.25}{\delta})} \implies
\frac{1.25}{\delta} < \exp{(\frac{\epsilon^2\sigma^2}{2\Delta_{2}(f)^2})} \\
\delta & > \frac{5}{4}\exp{(\frac{-\epsilon^2\sigma^2}{2\Delta_{2}(f)^2})}
\end{split}
\end{equation}
In our federated learning setting, $\it{f}$ is a neural network consisting of many layers of trainable parameters making computing $\Delta_{2}(f)$ intractable. However, without loss of generality, if we assume $\Delta_{2}(f)=1$, it is evident that increasing the level of $\sigma$ leads to a better bound for $\delta$. Following \citep{li2020multi}, we let $\sigma = \alpha\eta$, where $\eta$ is the standard deviation of the weight parameters of each layer in the neural network, effectively linking $\alpha$, an parameter adjustable for participating institutions, to the level of differential privacy protection. 


\subsection{Data preprocessing}
We processed and analyzed all of our WSI data at 20$\times$ magnification. Due to the lack of labeled ROIs and the intractable computational expense of deploying a convolutional neural network (CNN) directly to the whole spatial extent of each WSI, we utilize a form of weakly-supervised machine learning known as multiple instance learning (MIL). Under the MIL framework, each WSI is treated as a collection (bag) of smaller regions (instances), enabling the model to learn directly from the bag-level label (diagnosis or survival information) during training. The details of the MIL-inspired weakly-supervised learning algorithms we use are described in section 2.3 and 2.4. To construct the MIL bags, we utilize the CLAM  \citep{lu2020data} WSI processing pipeline to automatically segment the tissue regions in each WSI and divide them into M 256 $\times$ 256 image crops (instances), where M varies with the amount of tissue content in each slide. To overcome the computational challenges resulting from the enormous sizes of gigapixel WSI bags, each 256 $\times$ 256 RGB instance further undergoes dimensionality-reduction via a pretrained ResNet50 CNN encoder (truncated after the 3rd residual block for spatial average pooling), and is embedded as a 1024-dimensional feature vector for efficient training and inference. Accordingly, each WSI in the dataset is represented by a M $\times$ 1024 matrix tensor. For survival prediction, all WSIs corresponding each patient case are analyzed collectively, i.e., for a case with $N$ WSIs represented by individual bags of size $M_1, \cdots, M_N$ respectively, the bags are concatenated along the first dimension to form a single patient bag of dimensions $\sum_{j=1}^{N} M_j  \times 1024$.

\subsection{Weakly-supervised learning on WSIs}
We adopted an multiple instance learning-based framework for weakly-supervised classification and survival prediction and use it as the basis for performing federated learning on gigapixel WSIs. We begin by describing the weakly-supervised learning algorithms in the case of a single local model (no federated learning). Each model consists of a projection module $f_p$, an attention module $f_{attn}$, and a prediction layer $f_{pred}$. The projection module consists of sequential, trainable fully-connected layers that project the fixed feature embeddings obtained using a pretrained feature encoder into a more compact, feature space specific to histopathology images of the chosen disease model. Given the $j^{th}$ incoming WSI/patient bag of $M_j$ patch embeddings in the form of a $\mathbf{H}_j \in \mathbbm{R}^{M_j \times 1024}$ matrix tensor, for simplicity, we use a single linear layer $\mathbf{W}_{proj} \in \mathbbm{R}^{512 \times 1024}$ to project incoming patch-level embeddings into a 512-dimensional latent space. The attention module $f_{attn}$ uses attention-based pooling  \citep{ilse2018attention} to identify information rich patches/locations from the slides and aggregates their information into a single global representation for making a prediction at the bag level. We use the gated variant of the attention network architecture introduced by  \citep{ilse2018attention}. Accordingly, $f_{attn}$ consists of 3 fully-connected layers with weights $\mathbf{U}_{a}$, $\mathbf{V}_{a}$ and $\mathbf{W}_{a}$ and learns to assign an attention score to each patch embedding $\mathbf{h}_{j,m} \in \mathbbm{R}^{512}$ (each row entry in $\mathbf{H}_{j}$), indicating its contribution to the bag-level feature representation ${\mathbf{h}_{bag}}_{j} \in \mathbbm{R}^{512}$ , where $a_{j,m}$ represents the score for the $m^{th}$ patch and is given by:
\begin{equation}
a_{j,m}=\frac{\exp \left\{\mathbf{W}_{a}\left(\tanh \left(\mathbf{V}_{a} \mathbf{h}_{j,m}^{\top}\right) \odot \operatorname{sigm}\left(\mathbf{U}_{a} \mathbf{h}_{j,m}^{\top}\right)\right)\right\}}{\sum_{m=1}^{M_j} \exp \left\{\mathbf{W}_{a} \left(\tanh \left(\mathbf{V}_{a} \mathbf{h}_{j,m}^{\top}\right) \odot \operatorname{sigm}\left(\mathbf{U}_{a}\mathbf{h}_{j,m}^{\top}\right)\right)\right\}}
\end{equation}
Alternatively, the attention score vector for the whole bag is denoted by: $\mathbf{A}_{j}=f_{attn}(\mathbf{H}_{j})$.
Subsequently, the bag-level representation ${\mathbf{h}_{bag}}_{j}$ is calculated by using the predicted attention scores as weights for averaging all the feature embeddings in the bag as:
\begin{equation} \label{clsweightedaverage}
{\mathbf{h}_{bag}}_{j}=\textbf{Attn-pool}(\mathbf{A}_{j}, \mathbf{H}_{j})=\sum_{m=1}^{M_j} a_{j,m} \mathbf{h}_{j,m}
\end{equation} 
We used a 256-dimensional representation for the hidden layers in the attention network and apply Dropout with $p$=0.25 to these activations for regularization - namely, $\mathbf{U}_{a} \in \mathbbm{R}^{256 \times 512}$, $\mathbf{V}_{a} \in \mathbbm{R}^{256 \times 512}$ and $\mathbf{W}_{a} \in \mathbbm{R}^{1 \times 256}$. 
Lastly, the prediction layer $f_{pred}$ maps the bag representation 
${\mathbf{h}_{bag}}_{j}$ to predictions logits $\mathbf{s}_{j}$, using a different activation function and loss function for classification and survival prediction: $\mathbf{s}_{j}=f_{ pred}({\mathbf{h}_{bag}}_{j})$. The methodological details are described below. \\
\noindent\textbf{Weakly-supervised classification}
For weakly-supervised classification, we use the prediction layer $f_{pred}$ to predict the unnormalized class-probability logits $\mathbf{s}_{j}$, which are then supervised using the slide-level label $Y_j$ by applying the softmax activation and computing the standard cross-entropy loss. \\
\noindent\textbf{Weakly-supervised survival prediction} For weakly-supervised survival prediction using right-censored survival data, we consider discrete time intervals based on quantiles of event times for uncensored patients. More formally, we first consider the continuous time scale, where each labeled patient entry in the dataset, indexed by j, consists of a follow-up time $T_{j, cont} \in [0, \infty)$ and a binary censorship status $c_j$ where $c_j = 1$ indicates censorship (the event did not occur by the end of the follow-up period) while $c_j = 0$ indicates that the event occurred precisely at time $T_{j, cont}$. Next, we partition the continuous time scale $T_{cont}$ into R non-overlapping bins:  $[0, t_1), [t_1, t_2), ...., [t_{R-1}, \infty)$ and discretize $T_{j, cont}$ accordingly where:
\begin{equation}
T_{j, disc} = r \text{ iff } T_{j, cont} \in [t_r, t_{r+1})
\end{equation}
In our study, we use $R=4$ and choose $t_1, t_2 \text{ and } t_3$ based on the quartiles of the event times of uncensored patients.
For simplicity, from now on we refer to a patient's discrete survival time $T_{j, disc}$ simply as $T_j$ and to be consistent with the notation we used for classification, we refer to the ground truth label as $Y_j$. Given a patient's bag-level feature representation ${\mathbf{h}_{bag}}_{j}$ as calculated by the model, the prediction layer $f_{pred}$ is responsible for modeling the hazard function defined as:  
\begin{equation}
f_{hazard}(r \mid {\mathbf{h}_{bag}}_{j}) = P(T_j=r \mid T_j\geq r, {\mathbf{h}_{bag}}_{j})
\end{equation}
which relates to the survival function through:
\begin{equation}
\begin{split}
f_{surv}(r \mid {\mathbf{h}_{bag}}_{j}) & = P(T_j > r \mid  {\mathbf{h}_{bag}}_{j})\\
& = \prod_{u=1}^{r} (1 - f_{hazard}(u \mid {\mathbf{h}_{bag}}_{j}))
\end{split}
\end{equation}
Since we consider the label set $T_j \in \{0,\cdots,R-1\}$ to be the support of the hazard function, and $R = 4$ corresponding to quartiles of event times, $f_{pred}$ is a linear layer with weight parameters $\mathbf{W}_{pred} \in \mathbbm{R}^{4 \times 512}$.
Finally, given logits $\mathbf{s}_{j}=f_{ pred}({\mathbf{h}_{bag}}_{j})$, the sigmoid activation is applied to predict hazard distribution since it represents conditional probabilities, which are confined to positive real-values in the range of $[0, 1)$. 
For model optimization, we maximize the log likelihood function corresponding a discrete survival model \citep{tutz2016modeling}, which is written as:
\begin{equation}
\begin{aligned}
l = & \left(1-c_j\right) \cdot \log \left(\mathrm{P}\left(T_j=Y_j \mid {\mathbf{h}_{bag}}_{j} \right)\right) \\
& +c_j \cdot \log \left(f_{surv}\left(Y_j \mid {\mathbf{h}_{bag}}_{j}\right)\right)
\end{aligned}
\end{equation}
By rewriting $P(T_j = r \mid {\mathbf{h}_{bag}}_{j}) =
f_{hazard}(r \mid {\mathbf{h}_{bag}}_{j})f_{surv}(r \mid {\mathbf{h}_{bag}}_{j})$, the loss we minimize based on the log likelihood function \citep{zadeh2020bias} can be expressed as:
\begin{equation}
\begin{aligned}
L = - l & = - c_{j} \cdot \log \left(f_{surv}\left(Y_j \mid {\mathbf{h}_{bag}}_{j} \right)\right)\\ & - \left(1-c_j\right) \cdot \log \left(f_{surv}\left(Y_j-1 \mid {\mathbf{h}_{bag}}_{j} \right)\right)\\
& -\left(1-c_j\right) \cdot \log \left(f_{hazard}\left(Y_j \mid {\mathbf{h}_{bag}}_{j} \right)\right)
\end{aligned}
\end{equation}
During training, we additionally upweight the contribution of uncensored patient cases by minimizing a weighted sum of L and $L_{uncensored}$, which is defined by the terms:
\begin{equation}
\begin{aligned}
L_{uncensored} = & - \left(1-c_j\right) \cdot \log \left(f_{surv}\left(Y_j-1 \mid {\mathbf{h}_{bag}}_{j} \right)\right) \\
& -\left(1-c_j\right) \cdot \log \left(f_{hazard}\left(Y_j \mid {\mathbf{h}_{bag}}_{j} \right)\right)
\end{aligned}
\end{equation}
Accordingly, the loss we optimize for weakly-supervised survival prediction is:
\begin{equation}
L_{surv} = (1 - \beta) \cdot L + \beta \cdot L_{uncensored} 
\end{equation}

\subsection{Weakly-supervised federated learning with differential privacy}
For both classification and survival prediction, we train the models on each client server within a federated learning setup, where each model is trained locally and the weights of the model are collected each epoch and aggregated to update the central model. The central model then sends back the new weights to each client model. To preserve the differential privacy of the individual data located on each client server, we utilize a randomized mechanism, \textit{i.e.}, the Gaussian mechanism which we introduced in section 2.1. Hereby, our algorithm for collaboratively training server model and client models is shown in Algorithm \ref{alg1}.\\
In the proceeding section, we demonstrate the feasibility, adaptability and interpretability of attention-based multiple instance federated learning on four different computational pathology problems:  \textbf{A}) Breast Invasive Carcinoma (BRCA) subtyping \textbf{B}) Renal Cell Carcinoma (RCC) subtyping \textbf{C}) Clear Cell Renal Cell Carcinoma (CCRCC) survival prediction. 

\begin{figure*}
\includegraphics[width=\textwidth]{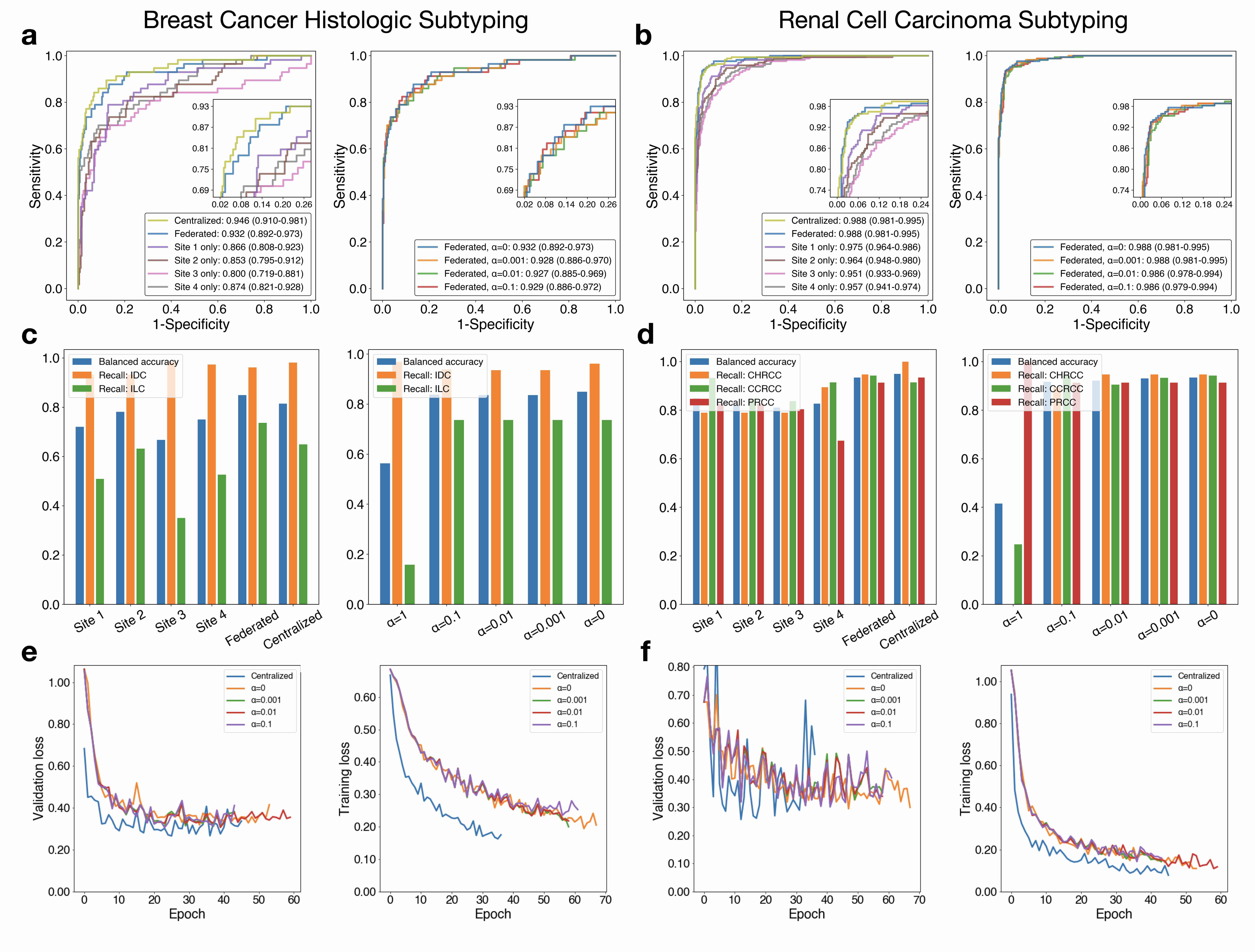}
\caption{\textbf{Performance and comparative analysis of privacy preserving weakly supervised federated learning on gigapixel whole slide images for histologic subtyping of breast carcinoma and renal cell carcinoma.} a-c, d-f The classification performance and loss curves of BRCA histologic subtyping and RCC histological subtyping. Top: ROC curves are generated on the test set for models trained using a centralized database, federated learning (with different levels of Gaussian random noise added during federated weight averaging) and using training data local to each institution individually. The AUC score along with its 95\% confidence interval (estimated using Delong's method) is reported for each experiment; micro-averaging is used for the multi-class classification of RCC subytping. Using multi-institutional data and federated learning, we achieved an AUC of between 0.927 and 0.932 on BRCA histologic subtyping and an AUC of between 0.986 and 0.988 on RCC histologic subtyping respectively. Middle: Balanced accuracy score and the sensitivity (recall) for each class (IDC: Invasive Ductal Carcinoma, ILC: Invasive Lobular Carcinoma for BRCA subtyping; CHRCC: Chromophobe Renal Cell Carcinoma, CCRCC: Clear Cell Renal Cell Carcinoma, PRCC: Papillary Renal Cell Carcinoma for RCC subtyping) is plotted for all experiments to assess model performance when accounting for class-imbalance in the respective test set. Bottom: For each experiment, the training loss and validation loss is monitored over each epoch before early stopping is triggered (see section 3.2). Federated training resulted in slower convergence on the validation set in both tasks.}
\end{figure*}

\renewcommand{\algorithmicrequire}{\textbf{Input:}}
\begin{algorithm}[ht!]
\caption{Privacy-preserving federated learning using attention-based multiple instance learning for multi-site histology-based classification and survival prediction}
\label{alg1}
\hspace*{\algorithmicindent} 
\\
\textbf{Input:} \\
I. WSI Data and weak annotation (\textit{e.g.} patient diagnosis or prognosis) scattered among B participating institutional sites: $(\mathbf{X},\mathbf{Y}) =  \left\{\left\{(\mathbf{X}_{1,j},Y_{1,j})\right\}, \cdots, \left\{(\mathbf{X}_{B,j},{Y}_{B,j})\right\}\right\}$, where $\left\{(\mathbf{X}_{i,j},{Y}_{i,j})\right\} = \left\{(\mathbf{X}_{i,1},{Y}_{i,1}), \cdots, (\mathbf{X}_{i,N_{i}},{Y}_{i,N_{i}}) \right\}$ represents the set of $N_{i}$ pairs of WSI data and corresponding label for training stored at site $i$ (in survival prediction, $\mathbf{X}_{i,j}$ is the set of all diagnostic WSIs for patient j whereas in classification, it is a single WSI). We use $(\mathbf{X}',\mathbf{Y}) =  \left\{\left\{(\mathbf{X}_{1,j}', {Y}_{1,j})\right\}, \cdots, \left\{(\mathbf{X}_{B,j}',{Y}_{B,j})\right\}\right\}$ to denote WSI data-label pair after patching and feature extraction via a pretrained CNN feature encoder. \\
II. Neural network models on local clients $\mathbf{\mathit{f_{local}}} = \left\{\mathit{f}_1,\cdots,\mathit{f}_B\right\}$ and global model $\mathit{f}_{global}$, stored on the central server. Each model $\mathit{f}_i$ consists of a projection module $f_{i, p}$, an attention module $f_{i, attn}$ and prediction layer $f_{i, pred}$. We denote the weights of the local models as $\left\{\mathbf{W}_{1},\cdots, \mathbf{W}_{B} \right\}$ and weights of the global model as $\mathbf{W}_{global}$. \\ 
III. Noise generator $M(\cdot)$, which generates Gaussian random noise $z\sim(0,\alpha\eta)$, where $\alpha$ denotes the noise level for and $\eta$ is the standard deviation of a neural network weight matrix. \\
IV. Number of training epochs or federated rounds, $K$. \\
V. Optimizers $\left\{opt_1(\cdot),\cdots, opt_B(\cdot)\right\}$, that update the model weights w.r.t a suitable loss metric $\it{L}$ using gradient descent.
\begin{algorithmic}[1]
\STATE {initialize local model weights $\left\{\mathbf{W}_{1}^{(0)},\cdots,\mathbf{W}_{N}^{(0)}\right\}$}
\FOR{$k=1$ to K}
\FOR{$i=1$ to B}
\FOR{$j=1$ to $N_{i}$} 
\STATE $\mathbf{H}_{i,j}=f_{i, proj}^{(k)}(\mathbf{X}_{i,j}')$ \\
$\mathbf{A}_{i,j}=f_{i, attn}^{(k)}(\mathbf{H}_{i,j})$\\
${\mathbf{h}_{bag}}_{i,j}=\textbf{Attn-pool}(\mathbf{A}_{i,j}, \mathbf{H}_{i,j})$\\
$\mathbf{s}_{i,j}=f_{i, pred}^{(k)}({\mathbf{h}_{bag}}_{i,j})$ \\
${\mathbf{W}_i}^{(k)}\leftarrow{opt}_i(\it{L}(\mathbf{s}_{i,j},Y_{i,j})))$ 
\ENDFOR
\ENDFOR
\STATE ${\mathbf{W}_{global}}^{(k)}\leftarrow \frac{1}{B}\sum_i({\mathbf{W}_i}^{(k)}+M({\mathbf{W}_i}^{(k)}))$ 
\FOR{$i=1$ to B}
\STATE ${\mathbf{W}_i}^{(k)}\leftarrow{\mathbf{W}_{global}}^{(k)}$
\ENDFOR
\ENDFOR
\RETURN{global model ${\mathit{f}_{global}}^{(K)}$}  
\end{algorithmic}
\end{algorithm}

\section{Experiments and Results}
\subsection{Dataset description} 
\noindent\textbf{Weakly-supervised classification.}
To evaluate the proposed federated learning framework for weakly-supervised classification in histopathology, we examined two clinical diagnostic tasks for two separate disease models, namely, Renal Cell Carcinoma (RCC) and Breast Invasive Carcinoma (BRCA). For both tasks, we used publicly available WSIs from the TCGA (The Cancer Genome Atlas) in addition to in-house data collected at the Brigham and Women's Hospital for model development and evaluation. In all cases, each gigapixel WSI is associated with a single ground truth slide-level diagnosis and no pixel or ROI-level annotation is available.\\
\textit{Breast cancer dataset.} For the first binary task of classifying primary Breast Invasive Carcinoma as either lobular or ductal morphologial subtypes,  1056 FFPE dianostic WSIs (211 lobular and 845 ductal) were retrieved from the TCGA BRCA (Breast Invasive Carcinoma) study and our in-house dataset consists of 1070 WSIs of primary breast cancer (158 lobular and 912 ductal). Accordingly, in total we used 2126 breast WSIs (369 lobular and 1757 ductal).
\\
\textit{Renal cell cancer dataset}
In the second task of multi-class classification of Renal Cell Carcinoma into clear cell (CCRCC), papillary cell (PRCC) and chromophobe cell (CHRCC) morphological subtypes, we collected 937 WSIs (519 CCRCC, 297 PRCC and 121 CHRCC) from the corresponding studies in TCGA and our in-house dataset consists of 247 WSIs of primary Renal cell carcinoma (184 CCRCC, 40 PRCC and 23 CHRCC). In total we used 1184 kidney WSIs (703 CCRCC, 337 PRCC and 144 CHRCC). \\

\begin{figure*}
\includegraphics[width=\textwidth]{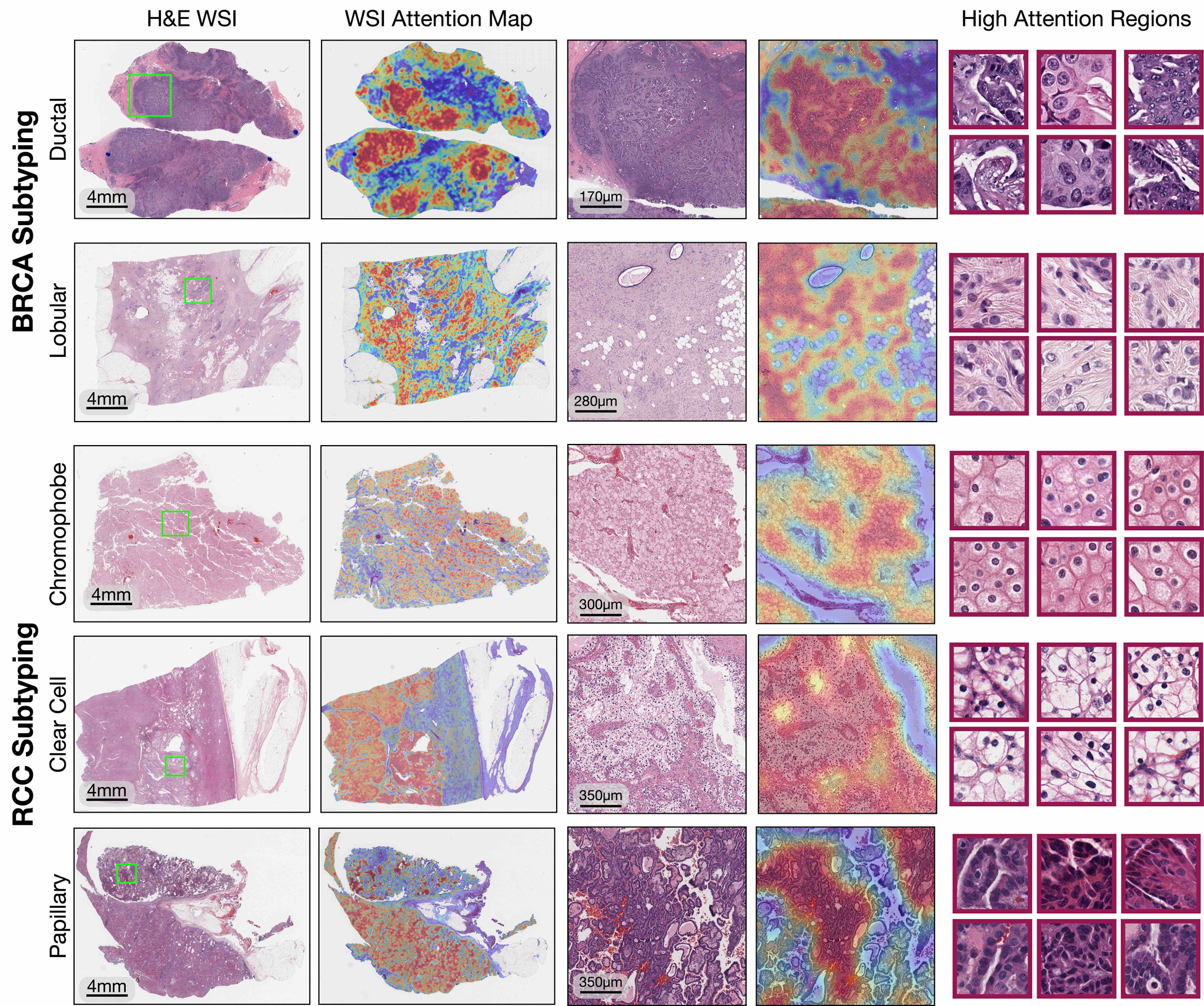}
\caption{\textbf{Interpretability and visualization for privacy preserving weakly-supervised federated classification}. In order to interpret and validate the morphological features learned by the model for RCC and BRCA histologic subtype classification, for randomly selected WSIs in the respective test set, the model trained with privacy-preserving federated learning ($\alpha=0.01$) is used to generate attention heatmaps using $256 \times 256$ sized tissue patches tiled at the 20 $\times$ magnification with a 90\% spatial overlap. For each WSI, the attention scores predicted for all patches in the slide are normalized to the range of $[0, 1]$ by converting them to percentiles. The normalized scores are then mapped to their respective spatial location in the slide. Finally, an RGB colormap is applied (red: high attention, blue: low attention), and the heatmap is overlaid on top of the original H\&E image for display. For BRCA, patches of the most highly attended regions (red border) exhibited well-known tumor morphology of invasive ductal carcinoma (round cells with varying degrees of polymorphism arranged in tubules, nests, or papillae) and invasive lobular carcinoma (round and signet-ring cells with intracellular lumina and targetoid cytoplasmic mucin arranged in a single-file or trabecular pattern). For RCC, highly attended regions exhibited well-known tumor morphology of chromophobe RCC (large, round to polygonal cells with abundant, finely-reticulated to granular cytoplasm and perinuclear halos), clear cell RCC (large, round to polygonal cells with clear cytoplasm and distinct, but delicate cell borders), and papillary RCC (round to cuboidal cells with prominent papillary or tubulopapillary architecture with fibrovascular cores). }
\end{figure*}

\noindent\textbf{Weakly-supervised survival prediction.} 
We also examined federated learning for weakly-supervised survival prediction based on histopathology. Specifically, for patients diagnosed with renal clear cell carcinoma, we used right-censored, overall survival data from the TCGA-KIRC available via the cbioportal. In total, 511 patient cases were retrieved from TCGA-KIRC. All diagnostic WSIs corresponding each patient case were used for analysis. 

\subsection{Experiments on multi-institutional WSI data}
In each of the two weakly-supervised classification tasks, we considered four distinct "institutional sites". These sites were identified by first naturally considering all in-house BWH data as one distinct institutional site. Then, for each TCGA cohort, we identified the tissue source site for each patient case. For the purpose of simulating federated learning across multiple institutions, we then randomly partitioned the set of unique tissue source sites into 3 non-overlapping, roughly equal-sized subsets, and grouped together the data corresponding to each subset of tissue source sites to serve as 3 distinct institutional sites. Similarly, for CCRCC survival prediction, we used 3 institutional sites created by randomly partitioning the tissue source sites that contributed to the TCGA-KIRC cohort. The details of these partitions are summarized 
below for each task (\textbf{Table 1,2 and 3}). 
\begin{table}[h!]
\centering
\caption{Partition for BRCA subtyping (number of WSIs)}
\begin{tabular}{llll}
\toprule
            & ILC & IDC  & Total \\
            \midrule
TCGA Site 1 & 56 & 155  & 211\\
TCGA Site 2 & 46 & 268 & 314 \\
TCGA Site 3 & 109 & 422 & 531 \\
BWH         & 158 & 912 & 1070 \\
\midrule
Total       & 369 & 1757 & 2126   \\
\bottomrule
\end{tabular}
\end{table}
\begin{table}[h!]
\centering
\caption{Partition for RCC subtyping (number of WSIs)}
\begin{tabular}{lllll}
\toprule
            & CCRCC & PRCC  & CHRCC & Total \\
\midrule
TCGA Site 1 & 108 & 120 & 39 & 267 \\
TCGA Site 2 & 78  & 100 & 31 & 209 \\
TCGA Site 3 & 333 & 77  & 51 & 461 \\
BWH         & 184 & 40  & 23 & 247 \\
\midrule
Total       & 703 & 337 & 144 & 1184 \\
\bottomrule
\end{tabular}
\end{table}
\begin{table}[h!]
\centering
\caption{Partition for CCRCC survival prediction (number of cases)}
\begin{tabular}{llll}
\toprule
            & Uncensored & Censored  & Total \\
            \midrule
TCGA Site 1 & 16 & 88  & 104\\
TCGA Site 2 & 27 & 49 & 76 \\
TCGA Site 3 & 128 & 203 & 331 \\
\midrule
Total       & 171 & 340 & 511   \\
\bottomrule
\end{tabular}
\end{table} \\
Once the institutional sites were identified, the dataset is then randomly partitioned into a training, validation and test set consisting of 70\%, 15\% and 15\% of patient cases respectively. For classification, given the class-imbalance nature of the datasets, within each institutional site, stratified sampling is used to ensure sufficient representation of minority classes across the training, validation and test set. Additionally, if a single patient case contains multiple diagnostic slides, all of them were drawn together into the same set when that patient is sampled. Similarly, for survival prediction, sampling is stratified based on both the discretized follow-up time (section 2.3) and the censorship status. \\
For each task, we used the model architecture and loss function as described in detail in section 2.3. To train each local model, we used the Adam optimizer with default hyperparamters, a learning rate of 2e-4 and L2 weight decay of 1e-5 for all experiments. For survival prediction, $\beta$, which controls how much the contribution of uncensored patients should be upweighted, was set to 0.15. Additionally, we monitored the validation loss each epoch and performed early stopping on the global model when it does not improve for 20 consecutive epochs (federated rounds), but only after it has been trained for at least 35 epochs. The model checkpoint with the lowest validation loss was then used to evaluation on the held-out test set. 
For each task, we investigated 3 scenarios: \textbf{1}) training on data from a single institution, \textbf{2}) training a single model by centralizing or pooling together all data (no federated learning) and \textbf{3}) training on data from all institutions using federated averaging, as described in section 2.4 and outlined in details in Algorithm 1. \\
For scenario \textbf{3}), we also studied changing the strength of Gaussian random noise added to local model weights during federated averaging, and its effect on the performance of the central model. As described in section 2.1, for each model $\it{f_i}$, we generated Gaussian noise $z_i \sim \mathcal{N}(0,\alpha\eta)$ where $\eta$ is the standard deviation of the weight parameters in each individual layer of the network and $\alpha$ controls the noise level. In our experiments, we varied $\alpha \in \{0,0.001,0.01,0.1,1.0\}$. In the next section, we present results demonstrating the effectiveness of weakly-supervised federated learning for both binary and multi-class classification, as well as survival prediction. 

\begin{figure*}[h!]
\includegraphics[width=\textwidth]{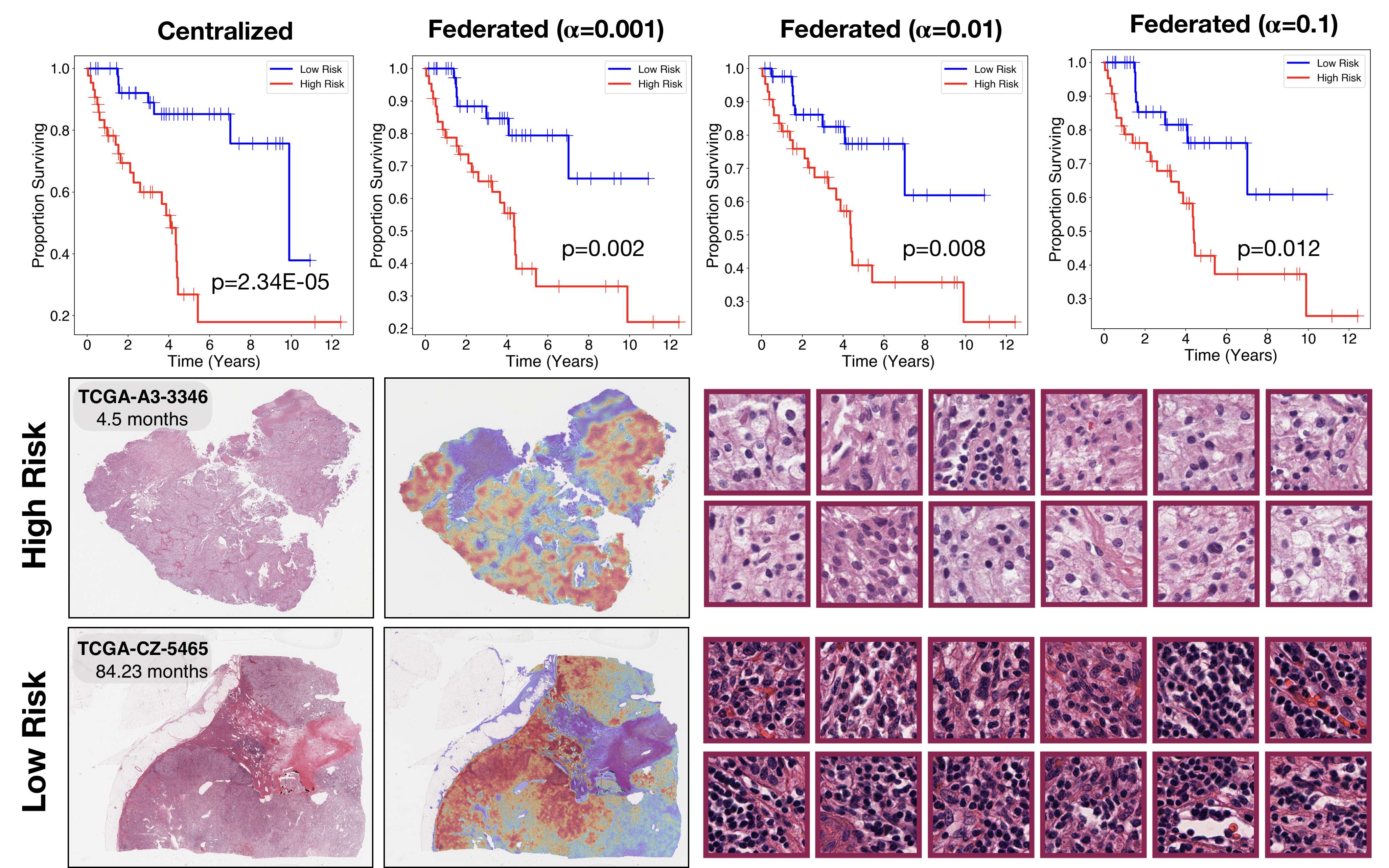}
\caption{\textbf{Patient stratification and interpretability for privacy preserving weakly-supervised federated survival prediction}. Patients in the test set were stratified into high risk and low risk groups using the median (50\% percentile) of the model's predicted risk score distribution as the cutoff and the log-rank test was used to assess the statistical significance between survival distributions of the resulting risk groups. Top: increasing $\alpha$ by over two orders of magnitude for stronger guarantees on differential privacy did not eliminate the model's ability to stratify patients into statistically significantly (p-value $<0.05$) different risk groups. Bottom: exemplars of Clear Cell Renal Cell Carcinoma patients predicted as high-risk and low-risk respectively by the model, the original H\&E (left), attention-based heatmap (center), and highest-attention patches (right). The highest attention patches for the high-risk case focus predominantly on the tumor cells themselves, while the highest attention patches for the low risk case focus predominantly on lymphocytes within the stroma and directly interfacing with tumor cells.}
\end{figure*}

\subsection{Experimental results}
We evaluated our proposed weakly-supervised, federating learning framework on both a multi-class and a binary  classification problem (\textbf{Figure 2, 3}, \textbf{Table 4 and 5}) as well as survival prediction (\textbf{Figure 4 and Table 6}) and demonstrated the feasibility of performing privacy preserving, federated learning on WSI data in all tasks. \\
In both BRCA subtyping (\textbf{Table 4}) and RCC subtyping (\textbf{Table 5}), the model performance is evaluated using a wide variety of classification metrics including the AUC of the ROC curve, mean average precision (mAP), classification error, F1 score, balanced accuracy (bAcc) and Cohen's $\kappa$ (micro-averaging is used to extend binary classification metrics to multi-class classification in the case of RCC subtyping). We found that the model performance benefited significantly from training on multi-institutional data using federated learning, compared to learning from data within a single institution. In fact, we found the models trained using federated learning to be generally competitive in performance even when compared to scenario \textbf{2}), where model is trained by first centralizing (sharing) all training data from each institution. This is true even when different levels of random noise are applied for privacy preservation. For $\alpha \in \{0, 0.001, 0.01, 0.1\}$, for BRCA subtyping, the test AUC (evaluated on n=321) ranged from 0.927 to 0.932 when using federated learning for different levels of random noise and for RCC subtyping, the micro-averaged test AUC (evaluated on n=170) ranged from 0.986 to 0.988. In addition to strong performance, in \textbf{Figure 3}, we also demonstrated that models trained using privacy-preserving federated learning can saliently localize regions of high diagnostic relevance and identify morphological features characteristic of each underlying tumor subtype.  
However, consistent with previous studies \citep{li2020multi}, we found that the model performance significantly deteriorated when $\alpha$ was set too high (\textit{e.g., $\alpha=1$}), showing that there is indeed a trade off between model performance and privacy protection. \\
For survival prediction, we evaluated the model performance using the c-Index, which measures the concordance in ranking patients by their assigned risk w.r.t. their ground truth survival time. Additionally, based on the predicted risk score for each patient in the test set, we performed hypothesis testing using the log-rank test to assess whether each model can stratify patients into distinct high risk and low risk groups (cutoff based on 50th percentile of the model's predicted risk scores) that resulted in statistically significantly different survival distributions (\textbf{Table 6}). When trained using data from a single institution, only 1 out of 3 institutions was able to yield a model that can stratify patients into distinct survival groups based on predicted risk scores. Notably, we observed that the model trained using data local to site 3 delivered performance comparable to that of centralized training and using federated learning. This can likely attributed to site 3 having a much larger local dataset (n=331) compared to the other 2 sites (n=104 and n=76 respectively). Similar dataset-size imbalance among different participating institutions frequently occurs in the real-world and is also reflected in the imbalanced distribution of patient cases among the original tissue source sites in the TCGA. In settings where the data at a single institution are insufficient (\textit{e.g.} site 1 and 2) in either size or diversity to yield a meaningful, generalizable model, soliciting data from collaborating institutions or other external sources may be necessary. On the other hand, we found that federated learning can overcome this challenge as all models trained in the federated framework (with the exception of when using $\alpha=1$) resulted in statistical significance (p-value $<0.05$) and produced c-Index values comparable with using centralized training. \\
Similar to classification, we visualized attention heatmaps over the entire WSI for low risk (long survival) and high risk (short survival) patients in order to interpret the regions and morphological features learned by the weakly-supervised model to be of high prognostic relevance. 

\begin{table*}[h!t]
\centering
\caption{BRCA subtyping performance on test set (n=321)}
\begin{tabular}{lcccccc}
\toprule
{} &           AUC (95\% CI) $\uparrow$ &  Error $\downarrow$ &   bAcc $\uparrow$ &     F1 $\uparrow$ &    mAP $\uparrow$ & Cohen's $\kappa$ $\uparrow$ \\
\midrule
Site 1 only        &  0.866 (0.808 - 0.923) &  0.143 &  0.720 &  0.558 &  0.667 &          0.473 \\
Site 2 only        &  0.853 (0.795 - 0.912) &  0.121 &  0.782 &  0.649 &  0.618 &          0.575 \\
Site 3 only        &  0.800 (0.719 - 0.881) &  0.128 &  0.668 &  0.494 &  0.605 &          0.434 \\
Site 4 only        &  0.874 (0.821 - 0.928) &  0.106 &  0.750 &  0.638 &  0.750 &          0.580 \\
\midrule
Centralized        &  0.946 (0.910 - 0.981) &  0.078 &  0.815 &  0.747 &  0.874 &          0.703 \\
\midrule
Federated, $\alpha=0.001$ &  0.928 (0.886 - 0.970) &  0.100 &  0.836 &  0.724 &  0.837 &          0.663 \\
Federated, $\alpha=0.01$  &  0.927 (0.885 - 0.969) &  0.100 &  0.836 &  0.724 &  0.834 &          0.663 \\
Federated, $\alpha=0.1$   &  0.929 (0.886 - 0.972) &  0.097 &  0.838 &  0.730 &  0.840 &          0.672 \\
Federated, $\alpha=1$     &  0.650 (0.569 - 0.732) &  0.174 &  0.564 &  0.243 &  0.341 &          0.176 \\
\bottomrule
\end{tabular}
\end{table*}

\begin{table*}
\centering
\caption{RCC subtyping test performance (n=170)}
\begin{tabular}{lcccccc}
\toprule

{} &           AUC (95\% CI) $\uparrow$ & Error $\downarrow$ & bAcc $\uparrow$ &     F1 $\uparrow$ &  mAP $\uparrow$ & Cohen's $\kappa$ $\uparrow$ \\
\midrule
Site 1 only        &  0.975 (0.964 - 0.986) &  0.112 &  0.850 &  0.888 &  0.955 &          0.787 \\
Site 2 only        &  0.964 (0.948 - 0.980) &  0.165 &  0.821 &  0.835 &  0.941 &          0.699 \\
Site 3 only        &  0.951 (0.933 - 0.969) &  0.176 &  0.811 &  0.824 &  0.917 &          0.683 \\
Site 4 only        &  0.957 (0.941 - 0.974) &  0.153 &  0.828 &  0.847 &  0.930 &          0.715 \\
\midrule
Centralized        &  0.988 (0.981 - 0.995) &  0.071 &  0.950 &  0.929 &  0.980 &          0.871 \\
\midrule
Federated, $\alpha=0.001$ &  0.988 (0.981 - 0.995) &  0.071 &  0.931 &  0.929 &  0.978 &          0.870 \\
Federated, $\alpha=0.01$  &  0.986 (0.978 - 0.994) &  0.088 &  0.922 &  0.912 &  0.975 &          0.839 \\
Federated, $\alpha=0.1$  &  0.986 (0.979 - 0.994) &  0.071 &  0.917 &  0.929 &  0.976 &          0.869 \\
Federated, $\alpha=1$    &  0.685 (0.638 - 0.732) &  0.576 &  0.416 &  0.424 &  0.539 &          0.129 \\
\bottomrule
\end{tabular}
\end{table*}

\begin{table}[h]
\centering
\caption{CCRCC survival prediction test performance (n=87)}
\begin{tabular}{lcc}
\toprule
{} & c-index $\uparrow$ &    p-value \\
\midrule
Site 1 only        &   0.532 &      0.115 \\
Site 2 only        &   0.536 &      0.141 \\
Site 3 only        &   0.713 &  1.937e-04 \\
\midrule
Centralized        &   0.715 &  2.336e-05 \\
\midrule
Federated, $\alpha=0.001$ &   0.713 &  1.847e-03 \\
Federated, $\alpha=0.01$  &   0.690 &  8.121e-03 \\
Federated, $\alpha=0.1$   &   0.700 &      0.012 \\
Federated, $\alpha=1$     &   0.539 &      0.095 \\
\bottomrule
\end{tabular}
\end{table}

\section{Conclusion}
In this work, we have demonstrated the feasibility and effectiveness of applying federated, attention-based weakly-supervised learning for general purpose classification and survival prediction on gigapixel whole slide images from different sites, without the need for institutions to directly share potentially sensitive patient data. Our proposed framework opens the possibility for multiple institutions to integrate their WSI datasets and train a more robust model that tends to generalize better on unseen data than models developed on data from a single institution, while also allowing participating institutions to preserve differential privacy via a randomized mechanism. Backed by a flexible and interpretable attention-based weakly-supervised learning framwork, we believe our federated learning framework has the clear potential to be applied to many important computational pathology tasks beyond what we have already shown in this study.
Decreasing barriers to cross-institutional collaborations in this way will be key to the future development of computational pathology tools. This is especially true in the case of rare diseases, where a single institution may not possess enough cases of a single entity to train an effective model on its own, due to a lack of diversity in morphology. These techniques may also be useful in situations where transferring large quantities of physical or digital slides may be impossible due to institutional or governmental regulations.  Models that give institutions greater control over their data while still achieving at or near state-of-the-art performance will be instrumental in progress towards democratized computational pathology.
\section*{Acknowledgements}

 This work was supported in part by internal funds from BWH Pathology, NIH National Institute of General Medical Sciences (NIGMS) R35GM138216A (to F.M.), Google Cloud Research Grant and the Nvidia GPU Grant Program. R.J.C. was additionally supported by the NSF Graduate Research Fellowship and NIH National Human Genome Research Institute (NHGRI) T32HG002295. 

\bibliography{ref.bib}
\end{document}